\documentclass{ws-procs975x65}

\def\beq{\begin{equation}}
\def\eeq{\end{equation}}

\begin{document}

\title{Topology Change in Spherical Gravitational Collapse}
\author{Sabbir A. Rahman$^*$}


\address{Department of Corporate Planning, Kingdom Economic and Energy Analysis Division,\\
Saudi Aramco, Dhahran 31311, Eastern Province, Kingdom of Saudi Arabia\\
$^*$E-mail: sarahman@alum.mit.edu}

\begin{abstract}
We argue that the formation of a Schwarzschild black hole via Datt-Oppenheimer-Snyder
type gravitational collapse must be accompanied by a change in topology upon
formation of the event horizon which physically separates matter in the interior
from that of the exterior. While it is true that collapsing matter crossing the
event horizon continues to fall towards the singularity of the Schwarzschild
interior, this region does not in fact contain the matter originally responsible
for the collapse. Rather, the latter occupies a distinct internal
spacetime region with its own independent evolution.
The existence of this additional component of the simplest black hole
has a number of profound implications - Schwarzschild black holes are stable
with constant mass; they each contain a self-contained mini-universe within their
event horizons; and they live within a spacetime that is inherently double-sheeted.
\end{abstract}


\bodymatter


\section{Three Issues with the Standard Picture of Spherical Collapse}

A survey of the literature relating to (classical, idealised) Datt-Oppenheimer-Snyder type spherical gravitational collapse\cite{Oppenheimer39,Datt38} indicates that there are a number of issues which do not yet appear to have been adequately addressed or resolved, including specifically (i) the non-spherical geometry\cite{Brillouin23,Bonnor92} of the Schwarzschild interior, (ii) the location of the mass sources, and (iii) the presence of the central singularity.

\subsection{Resolving the interior geometry of the collapsing matter}
In the standard description of spherical gravitational collapse\cite{Finkelstein58}, the metric in the interior of the collapsing matter is acknowledged {\it not} to be the Schwarzschild interior metric prior to formation of the black hole, but once the black hole has formed, its metric transforms into the Schwarzschild interior complete with singularity. After this transformation all the matter already inside, or subsequently falling into, the black hole then collapses into the singularity. Although this account of spherical collapse has by now become well established, the description of what happens to the collapsing matter interior to the critical surface cannot in fact be correct.

One way to see why the standard picture is problematic is to consider the case of a collapsing thick spherical shell of total mass $m$. As the leading (i.e innermost) edge of the shell - and subsequently all of the mass except for the outermost edge of the shell - passes across the Schwarzschild radius $r=2m$, the mass associated with these inner subshells will change the metric continously inside $r=2m$, and, because the critical mass has not yet been reached inside the Schwarzschild radius, this deformed metric will just be an ordinary non-black hole interior metric. This has to be the case because no black hole forms until the entire thick shell has crossed $r=2m$, and indeed prior to that happening, there is always in principle the possibility that an outward thrust can be applied to prevent the outermost subshell from collapsing past $r=2m$, thus preventing the formation of a black hole. 

Once the outermost subshell has crossed $r=2m$ the formation of the black hole is confirmed, and it is therefore this outermost subshell that must trace out the new black hole interior metric that gives rise to the formation of the singularity, so that any matter exterior to the subshell which subsequently falls in will hit the singularity. On the other hand, the inner subshells, which have long passed $r=2m$ will be completely oblivious to this, and will still be living in a non-black hole interior. Being of infinitesimal mass, the outermost subshell can only generate continuous local changes to the metric at its current position, and it cannot somehow effect the discontinuous change that would be required at the location of the inner subshells to transform the non-black hole metric into the black hole one that would cause the inner subshells to fall into the singularity that would eventually form.

The only reasonable way to explain the independent coexistence of these two distinct interior metrics is if there is a topological bifurcation that takes place at the Schwarzschild surface upon formation of the black hole whereby the original manifold splits into two - one containing the bulk of the infalling shell (i.e. the entire mass $m$) endowed with a non-black hole interior metric, and the other containing all matter outside of the shell which happens to cross $r=2m$ that corresponds to the new black hole interior region. The matter originally responsible for the collapse therefore becomes physically separated from the rest of the universe (i.e. the Schwarzschild manifold) - analogous to the formation of a soap bubble - and lives in its own independent universe connected to the Schwarzschild manifold only at the Schwarzschild surface $r=2m$ where the topology change occurred. The likelihood of the occurrence of such critical phenomena at the event horizon had actually already been hinted at by Finkelstein.\cite{Finkelstein58} The same considerations mentioned above also apply directly to more general spherically symmetric configurations of collapsing matter, and it is highly probable that the formation of {\em any} black hole via gravitational collapse will be accompanied by a corresponding change in topology.

\begin{figure}
\begin{center}
\includegraphics[scale=0.45]{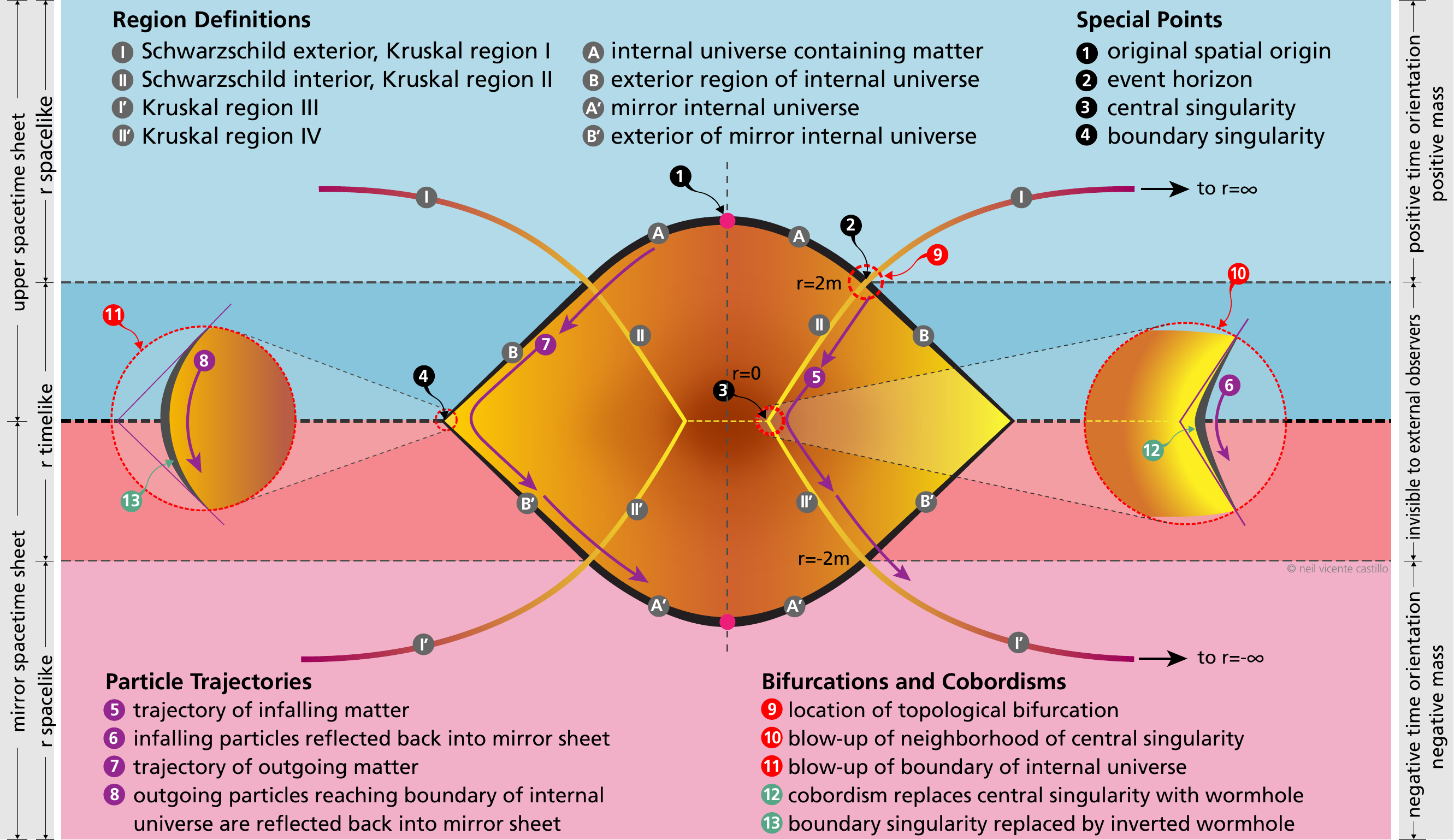}
\caption{The topology of the Schwarzschild black hole complete with internal universe.}
\end{center}
\end{figure}

The alternative description of spherical gravitational collapse given above satisfactorily addresses the objections of Brillouin\cite{Brillouin23} and Bonnor\cite{Bonnor92} in that the manifold containing the source matter retains its spherical geometry while still resulting in the formation of the Schwarzschild interior. It also resolves the second issue mentioned regarding the location of the mass sources in a natural way.

\subsection{Resolving the location of the mass sources}

It has generally been taken for granted  that the mass of the Schwarzschild black hole must somehow be associated with the central singularity. However a very careful analysis by Lynden-Bell and Katz\cite{LyndenBell90} has shown that this assumption is incorrect and that there is no matter to be found there. The discussion above makes clear that the mass source is actually associated with the event horizon, which is in accord with previous analyses carried out independently by a number of authors.\cite{McCrea64,Bel69}

Lynden-Bell and Katz also show in the same paper that a particle entering the Schwarzschild interior and reaching the central singularity emerges on the second `white hole' Kruskal sheet\cite{Finkelstein58,Kruskal60} travelling backwards in time.\footnote{This implies that the singularity acts as a kind of reflecting barrier preventing further collapse that might otherwise cause an increase in mass of the black hole.} In particular, they emphasise that this result is unique, and argue that since the singularity is crossed in an infinitesimal amount of time its presence may not be a reason for particular concern. There have been a number of earlier proposals that the black hole and white hole solutions can be sewn in this way\cite{Finkelstein58,Israel66,Ellis73}, and they have in common that they all give rise to a picture in which the Schwarzschild black hole takes on the appearance of a wormhole\footnote{Identifying such wormholes with elementary particles can explain the origin of electrodynamics as a purely emergent theory from classical gravity,\cite{Rahman09} with the associated electric field lines exerting an outward pressure that prevent the wormhole from pinching shut\cite{Graves60,Fuller62} and a singularity forming.} straddling two spacetime sheets, with the implication that the universe itself is inherently double-sheeted.\cite{Sakharov70,Chardin08,Rahman09} That is to say, the universe consists of two superimposed four-dimensional spacetime sheets for a total of eight dimensions - just large enough to accommodate the standard model gauge group.\cite{Trayling01}

\subsection{Resolving the issue of the central singularity}

Genuine singularities are unlikely to exist in nature and Yodzis' work on Lorentz cobordisms\cite{Yodzis72} shows that the Schwarzschild singularity can be consistently replaced by a wormhole using a simple cutting and sewing procedure. Ellis\cite{Ellis73} also provides an explicit construction of such a cobordism. According to Geroch's theorem,\cite{Geroch67} a change in topology must be accompanied by a change in the direction of time, and this is in fact precisely what occurs here, and so the picture we portray is consistent in that sense, even though it entails loss of causality. As pointed out by Israel,\cite{Israel67} the loss of causality associated with the full Kruskal extension need not be something to fear, and it could in fact turn out to be essential to a proper understanding of the physical foundations of both quantum theory and electrodynamics.\cite{Rahman09}

Turning our attention to the internal spherical region containing the matter responsible for formation of the black hole, it is not clear whether the boundary is unconstrained so that the pressure from particles in the interior will cause the region to expand relatively freely, or whether the boundary is constrained resulting in an oscillatory universe. However it seems fairly certain that because of the pinching that takes place at the Schwarzschild surface, particles reaching the outer boundary will be reflected `backwards in time' onto a second spherically symmetric sheet, analogous to the second sheet of the Kruskal extension, after first passing through a second exterior region - in fact a kind of inverted wormhole - with temporal and radial directions swapped in analogy with the Schwarzschild interior. The resulting extended topological structure of the Schwarzschild black hole, which includes the additional regions discussed above, is depicted in Figure 1.

\section{Some Further Consequences of the Proposed Topology Change}

In each of these scenarios, the following conclusions hold: First of all, the matter originally responsible for the black hole will continue to evolve in its own internal mini-universe independent of the rest of the solution. This means that further gravitational collapse can occur, leading to a hierarchy of black holes or mini universes located within larger black holes, all arranged within a fractal-like structure.

Because all matter falling through the event horizon is reflected into the second spacetime sheet, spacetime must be double-sheeted, with the orientation of time (and the sign of mass) on either sheet reversed relative to that on the other. Moreover, the decoupling of the internal universe from the exterior means that black holes will be stable and of fixed mass. As a further corollary, black holes will be gravitational dipoles, and if it happens to be the case that all elementary particles are black holes, as a number of models have proposed, this would mean that the vacuum, which is replete with neutrinos for example, is gravitationally polarisable, leading to an explanation for the existence of modified Newtonian dynamics.\cite{Blanchet06,Hu13}.

As a final remark, we agree with Ellis\cite{Ellis13} that while there is information loss upon formation of a black hole, there is no information paradox.

\section*{Acknowledgments}

I would like to thank King Fahd University of Petroleum \& Minerals for granting me access to their facilities during the preparation of this work, and also Rashid Kidwai and Muhammad Akmal for their kind support throughout.

\end{document}